\documentclass{article}

\usepackage{arxiv}

\usepackage[utf8]{inputenc} 
\usepackage[T1]{fontenc}    
\usepackage{hyperref}       
\usepackage{url}            
\usepackage{booktabs}       
\usepackage{amsfonts}       
\usepackage{amsmath}
\usepackage{nicefrac}       
\usepackage{microtype}      
\usepackage{lipsum}
\usepackage{graphicx}
\graphicspath{ {./images/} }

\title{Quantitative phase imaging of single particles from a cryoEM micrograph}

\author{
  Apoorv Pant \\
  Department of Physics\\
  Indian Institute of Technology Delhi\\
  New Delhi 110016 India\\
   \And
  Manidipa Banerjee \\
  Kusuma School of Biological Sciences\\
  Indian Institute of Technology Delhi\\
  New Delhi 110016 India \\
  \And
  Kedar Khare \\
  Department of Physics\\
  Indian Institute of Technology Delhi\\
  New Delhi 110016 India\\
  \texttt{kedark@physics.iitd.ac.in} \\
}

\begin{document}
\maketitle
\begin{abstract}
We show that de-focused single particle images recorded using a cryo-electron microscope (cryoEM) system may be processed like a Fresnel zone in-line hologram to obtain physically meaningful quantitative phase maps associated with individual particles. In particular, a region-of-interest (ROI) of the de-focused image surrounding a particle can be numerically back-propagated, in order to determine accurate de-focus information based on the sparsity-of-gradient merit function. Further with the knowledge of de-focus information, an iterative Fresnel zone phase retrieval algorithm using image sparsity constraints can accurately estimate the quantitative phase information associated with a single particle. The proposed methodology which can correct for both de-focus and spherical aberrations is a deviation from the image processing chain currently used in single particle cryoEM reconstructions. Our illustrations as presented here suggest that the phase retrieval approach applies uniformly to de-focused image data recorded using the traditional CCD detectors as well as the newer direct electron detectors.
\end{abstract}
\keywords{CryoEM, Fresnel zone phase retrieval, single-particle imaging, inline holography, quantitative phase imaging, image sparsity }

\section{Introduction}
The developments in the area of cryo-electron microscopy (cryoEM) over the last decade have revolutionized structural biology by enabling 3D reconstruction of bio-molecules with resolution approaching that offered by X-ray crystallography \cite{KW2014}. A key advantage of cryoEM is that it is not required to crystalize the bio-molecules \cite{OS2011}, thereby reducing the time required for structure determination to a matter of weeks or months. The recent developments in cryoEM involving introduction of hardware components like phase plates \cite{DN2010}, direct electron detectors \cite{BAM2012} and spherical aberration correctors \cite{MULLER2011} have helped in significantly improving the achievable 3D resolution and has thus generated considerable interest in cryoEM imaging worldwide \cite{Yip2020,Nakane2020}.  

Single-particle reconstruction from cryoEM images is a challenging task due to the relatively high level of noise. The noise level in the recorded cryoEM images cannot be reduced due to the inherent limitation on allowable electron dosage. By now a well-developed image processing chain exists in the form of open source software tools such as EMAN \cite{TANG2007}, RELION \cite{SCHERES2012a,SCHERES2012b} and many others that are currently being used routinely by a number of structural biologists worldwide. The image processing chain involves steps such as particle picking, contrast transfer function (CTF) estimation, Wiener filtering or other methods for CTF correction, class averaging, and particle alignment followed by 3D reconstruction \cite{JF2006}. Considerable efforts have gone into developing algorithms that have gone into these software tools that are able to provide impressive high quality image reconstructions to the cryoEM users from highly noisy micrographs. 

De-focused cryoEM micrographs are known to represent in-line Fresnel zone electron wave holograms \cite{GABOR1948, WADE1992} (this is how Dennis Gabor initiated his work in holography) formed by the interference of the scattered object wave and the unscattered wave.  
Recovery of complex-valued exit wave in transmission electron microscopy based on a series of de-focused micrographs using an approach that combines the transport of intensity equation and iterative algorithms has been demonstrated \cite{KOCH201469}. 
This in-line holography aspect of the cryoEM data however appears to have been somewhat lost in how the cryoEM reconstruction methodology has evolved traditionally. The basic framework regarding the definition of CTF in cryoEM is indeed based on assumption that the particles being imaged are nearly pure phase objects and the various steps in reconstruction methodology do utilize ideas from Fourier optics. However the reconstructed image by current cryoEM algorithms cannot be claimed to provide physically meaningful quantitative phase information about the individual macro-molecular particle images. Over the same time period when cryoEM reconstruction methodologies were being developed, Fourier and Fresnel zone phase retrieval ideas in optics literature have made tremendous progress that has somehow not made its way into the cryoEM algorithms to the best of our knowledge. Our tests with standard cryoEM data reveal that in-line holography methodologies can detect the focus plane by simple Fresnel back-propagation of the recorded local ROI of a single particle. The de-focus value is found to vary from particle to particle with a mean nearly equal to the nominal de-focus value based on the standard CTF estimation methods. Further, with the knowledge of accurate de-focus value, the variant of a sparsity assisted Fresnel zone phase retrieval algorithm   \cite{CG2019} can estimate quantitative phase information associated with a single particle by using only the local ROI image information. The methodology does not suffer from the usual problems associated with CTF estimation from noisy data and also with the zeros of CTF that make it difficult to perform Wiener filtering. Further we find that the iterative Fresnel-zone phase retrieval provides a complex-valued exit wave (transmitted field from a single particle) which is consistent with the recorded intensity data. The complex-valued exit wave field can thus be associated with physically meaningful quantitative phase information for an individual particle. The illustrations shown in this paper represent our initial work and we believe that this line of thought may lead to significant research activity in cryoEM reconstruction in future.

The paper is organized as follows. In Section 2 we test a focus plane detection criterion using local ROI information for single particles from a de-focused cryoEM micrograph frame. The focus plane detection methodology is illustrated for multiple particles in a single micrograph. In Section 3, we lay out a noise robust iterative framework for Fresnel zone phase recovery from de-focused images of single particles and show phase recoveries for a micrograph from the standard KLH (Keyhole Limpet Hemocyanin) dataset \cite{ZHU2004}. Section 4 presents iterative phase recoveries using a micrograph from the Apoferritin dataset EMPIAR-10146 that has been recorded the newer direct electron detector to illustrate that the proposed methodology for quantitative phase imaging can be applied uniformly to generic cryoEM data. In Section 5 we will discuss our conclusions and thoughts on future directions. 

\section{Test of focus detection metric for cryoEM data}
In this and the later Sections we will use the standard Fourier optics model for detected cryoEM images. In particular, for an electron wave-function $\psi_0(x,y)$ incident on a sample located in the $z = 0$ plane, the exit wave transmitted through the sample is given by:
\begin{equation}\label{eq:exitwave}
    g(x,y) = \psi_0(x,y) \, t(x,y).
\end{equation}
Here $t(x,y)$ represents the complex valued transmission function of a single particle. The function $t(x,y)$ is usually assumed to represent a weak-phase object of the form:
\begin{equation}\label{eq:transmission}
    t(x,y) = A(x,y) \exp(i \theta(x,y)) \approx A_0 [1 + i \theta(x,y)].
\end{equation}
The recorded cryoEM images have two main aberrations in the form of de-focus $\Delta z$ and the spherical aberration described by the coefficient $C_s$. The spherical aberration is present inherently in the system whereas the de-focus is introduced while recording an image in order to introduce contrast in the image. The coherent transfer function for the imaging system may therefore be described in terms of spatial frequency coordinates $(f_x, f_y)$ as \cite{OS2011}:
\begin{equation}\label{eq:system_model}
    H(f_x, f_y) = H_0(f_x,f_y) \exp[-i \pi \lambda \Delta z \rho^2 + i (\pi/2) C_s \lambda^3 \rho^4],
\end{equation}
with $\rho = \sqrt{f_x^2 + f_y^2}$. In the above relation, $H_0(f_x,f_y)$ is a slowly varying envelope function, which for the present work will be assumed to be a constant, and $\lambda$ is the electron wavelength. The recorded image irradiance may therefore be described as per the linear system model of imaging as:
\begin{equation}
    I(x,y) = |\mathcal{F}^{-1} \{ \mathcal{F}[g(x,y)] \;\; H(f_x,f_y)   \;\} |^2.
\end{equation}
Here the notation $\mathcal{F}\{...\}$ represents the two-dimensional Fourier transform operation. The image formation hardware chain in cryoEM systems is similar to a 4F system in Fourier Optics literature \cite{goodman2005}. Due to the form of the transmission function as in Eq. (\ref{eq:transmission}), the recorded image $I(x,y)$ may be considered to be an in-line electron wave hologram \cite{WADE1992}. 

The conventional method of estimating the de-focus of a micrograph involves modeling the Contrast transfer function (CTF) of the system and comparing this model with the position of Thon rings in the micrograph's 2D-Fourier transform. The system CTF is given by:
\begin{equation}\label{eq:CTF}
    CTF(\mathbf{\rho})=w\cos{\mathbf{\chi}(\mathbf{\rho})}+\sqrt{1-w^2}\sin{\mathbf{\chi}(\mathbf{\rho})},
\end{equation}
where $w$ is the fraction of the total contrast attributed to sample absorption (amplitude contrast) and $\mathbf{\chi}(\mathbf{\rho})$ is the frequency dependent phase-shift given by:
\begin{equation}\label{eq:CTF2}
    \mathbf{\chi}(\mathbf{\rho})=-\pi\lambda\Delta z \mathbf{\rho}^2+\frac{\pi}{2}C_s\lambda^3\mathbf{\rho}^4.
\end{equation}
As we can observe from (\ref{eq:CTF2}), the phase shift $\mathbf{\chi}(\rho)$ depends on the electron wavelength $\lambda$, the de-focus (to be estimated) and the spherical aberration coefficient $C_s$ which is generally known. The de-focus for the micrograph is estimated using a least-square fitting of the model CTF and the 2D Fourier transform of the micrograph within a range of spatial frequencies. The amplitude contrast $w$ is generally set to a small value (such as 0.07 or 0.1) and not explicitly measured for each measurement. There are several open-source software available for CTF estimation such as CTFFIND4 \cite{RG2015} and GCTF \cite{ZHANG2016}, that are based on above the principle. These software tools estimate de-focus of the micrograph as a whole and not for individual particles. Although recent versions of reconstruction software implement CTF refinement to compensate for  variations in particle-to-particle de-focus, this is done at a later stage of the reconstruction pipeline when a 3D density map is already available \cite{Relion3}.

The discussion above is part of the standard description of cryoEM systems. We however deviate from the standard image processing chain. While the usual processing methodology centers around aligning and averaging a number of individual particles, we focus on retrieving quantitative phase information just using cropped local ROI surrounding a single particle. While the high level of noise in the data is certainly a challenge in this task, we observe that meaningful information may be recovered out of a local single particle ROI.  

\begin{figure}[tb]
    \centering
    \includegraphics[width = 0.5\textwidth]{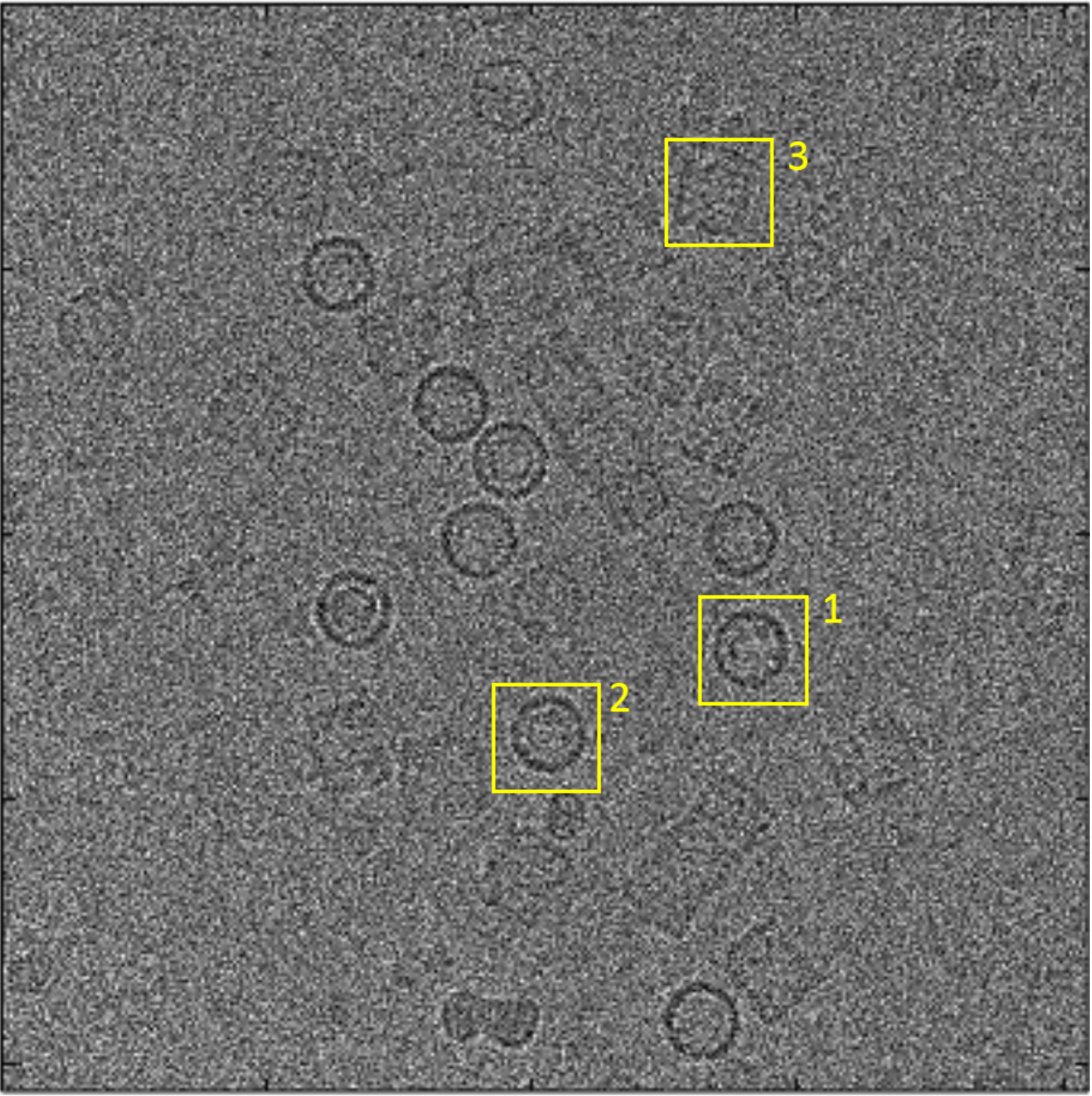}
    \caption{A typical cryoEM image from the KLH dataset recorded on a Phillips CM 200 system with a nominal de-focus value of around $2.7 \mu$m estimated using CTFFIND4. The pixel size is $2.2$ Angstrom and the image has $2048 \times 2048$ pixels. The boxed regions show the ROIs used for phase recovery.}
    \label{fig1}
\end{figure}

A typical cryoEM micrograph is shown in Fig. \ref{fig1}. This micrograph has been taken from the standard KLH data-set recorded using the Phillips CM 200 system operated at 120 kV. The de-focus value for this specific micrograph is around $2.7 \, \mu$m (estimated using CTFFIND4) and the pixel size after accounting for magnification is $2.2$ Angstrom. This micrograph has been categorized as a part of the farther from focus (FFF) data with a nominal de-focus value of $3 \mu$m. 
Our first task is to determine the accurate de-focus distance $\Delta z$ for an individual particle. For this purpose we take a $256 \times 256$ pixel ROI centered on an individual particle and perform the back-propagation operation (including the effect of spherical aberration) that is a standard procedure for handling an in-line hologram in optical imaging. The extent of blurring in the single particle due to de-focus and spherical aberrations is visible in the image and the ROI size has been selected to be much larger than that. The ROI image is cropped out of the recorded data frame $I(x,y)$ as in Fig. \ref{fig1} and will be denoted as $I_{ROI}(x,y)$. The ROI image is processed with the filter $H^{\ast}(f_x,f_y)$ as follows:
\begin{equation}\label{eq:backprop}
    q(x,y; z) = \mathcal{F}^{-1} \{ \mathcal{F}[I_{ROI}(x,y)] \; \exp[ i \pi \lambda z \rho^2 - i (\pi/2) C_s \lambda^3 \rho^4].
\end{equation}
The filtering operation above is performed in small steps of $\delta z = 5$ nm in $z$ over the range $z_1 = 2 \, \mu$m to $z_2 = 4 \, \mu$m. Since the above operation is implemented in computer using the FFT algorithm, the ROI image $I_{ROI}(x,y)$ is first even- symmetrized to avoid boundary artifacts. After the back-propagation computation, the top left quadrant of the result is used for further processing. As shown by Volkov et al in the context of the transport of intensity equation, this symmetrization operation conserves the total energy in the field within the computational boundary \cite{VOLKOV2002}. We note that $q(x,y; z)$ is a complex-valued function. In order to find the appropriate numerical value of de-focus $\Delta z$, we evaluate the Sparsity of Gradient (SoG) merit function \cite{Zhang17, Tamamitsu2018} (also known as the Tamura coefficient) given by:
\begin{equation}\label{eq:merit}
    M(z) = \sqrt{\sigma_u / \langle u \rangle},
\end{equation}
where $u(x,y; z)$ is the gradient magnitude image:
\begin{equation}
    u(x,y;z) = \sqrt{|\nabla_x q(x,y;z)|^2 + |\nabla_y q(x,y;z)|^2}.
\end{equation}
In Eq. (\ref{eq:merit}), $\sigma_u$ and $\langle u \rangle$ stand for the standard deviation and mean of the numerical pixel values in $u(x,y;z)$. The numerical value of $z$ for which the merit function $M(z)$ is maximum over the range of $z_1 = 2 \mu$m to $z_2 = 4 \mu$m is then treated as the de-focus $\Delta z$ for that particular particle.  Sparsity of gradient is known to be a robust criterion for focus plane detection in the context of in-line holograms in optics literature. It has been shown to yield accurate de-focus distance estimate even in cases when the weak phase approximation for the transmission function $t(x,y)$ does not hold. In Fig. 2 we show the result of the back-propagation process described above for a cropped ROI centered on the particle marked as ``1'' in Fig. \ref{fig1}. The raw ROI image is shown in Fig. 2(a). The corresponding plot of the merit function $M(z)$ is shown in Fig. 2(b) which shows that the de-focus value for the particle $\Delta z$ is given by $3.67\, \mu$m. Finally the amplitude and phase of the back-propagated functions $q(x,y; z)$ for $z = \Delta z$ (obtained as per maximum of the merit function) are shown in Fig. 2(c), (d) respectively. The phase map of $q(x,y;z=\Delta z)$ clearly shows features of the single particle structure, however, this phase map is not physically meaningful as it is corrupted by the extra terms (the dc and twin terms) that are present in the in-line hologram replay. 
\begin{figure}[tbp]
    \centering
    \includegraphics[width = 0.5\textwidth]{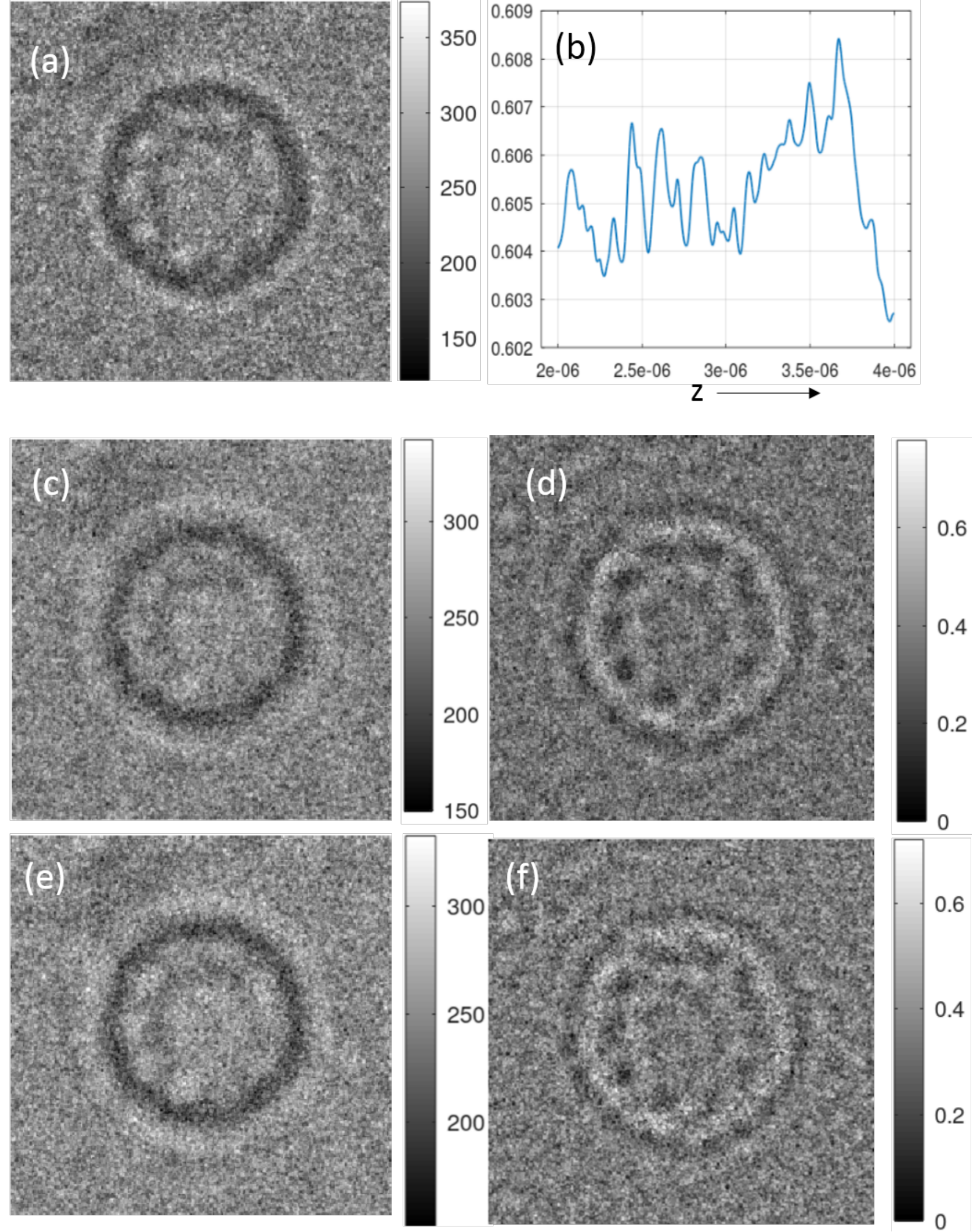}
    \caption{Result of applying the back-propagatipon process as in Eq. (\ref{eq:backprop}) to the ROI image cropped from the micrograph in Fig. \ref{fig1}. (a) cropped ROI image, (b) plot of merit function $M(z)$ over distance $z = 2 \mu$m -$4 \mu$m in steps of $5$nm. The peak position of the merit function corresponds to the appropriate $\Delta z$, (c) Amplitude and (d) phase of the back-propagated complex-valued function $q(x,y,z=\Delta z)$ with $\Delta z = 3.67 \mu$m, which is the location of the peak of the merit function $M(z)$ as shown in (b). The amplitude and phase show the structure of the single particle, however they are corrupted by the extra terms (dc and twin) in the in-line hologram. (e) Amplitude and (f) phase of the back-propagated function $q(x,y,z)$ when $z = 2.7 \mu$m is used as per the estimated de-focus of the whole micrograph.} 
    \label{fig2}
\end{figure}
We note that the de-focus value $\Delta z$ for the individual particles differs from the nominal de-focus value of $2.7 \mu$m that was estimated using CTFFIND4 \cite{RG2015} using the whole micrograph. For $20$ different particles in the micrograph in Fig. 1, the de-focus values $\Delta z$ as obtained using this approach have mean and standard deviation equal to $2.9 \pm 0.7 \mu$m respectively. It it important to note that only the local ROI image of a particle has been used in the de-focus estimation for individual particles and the process did not involve matching the noisy Thon rings associated with the micrograph to the CTF model. We additionally show the amplitude and phase of the back-propagated the ROI image $q(x,y;z)$ in Fig. \ref{fig2}(e),(f) when the value of $z = 2.7 \mu$m (as known for the whole micrograph) is used. We note that while the amplitude and phase are still not physically meaningful, the contrast in the phase image as in Fig. \ref{fig2}(f) is slightly lower compared to that in Fig. \ref{fig2}(d) obtained using de-focus estimated using peak of the merit function. We note that the range of numerical values taken by the merit function $M(z)$ is small, however, based on comparison of the phase maps in Fig. \ref{fig2}(d),(f), the peak seen at $z = 3.67 \mu$m may have significance as we will see again later in this paper. While sparsity of gradient merit function has been shown to provide robust de-focus estimates in optical in-line holography, the accuracy of this approach for cryoEM particle images needs further study, since the noise in cryoEM micrographs is much higher. A detailed comparative study of this defocus estimation method with the conventional CTF estimation procedure needs to be performed in future.

\section{Fresnel-zone phase retrieval for individual single-particles}
In the previous section we showed that applying the sparsity of gradient merit function allows us to determine the de-focus $\Delta z$ for an individual particle using the local ROI information alone. Our next task is to determine the complex valued exit wave $g(x,y)$ as defined in Eq. (\ref{eq:exitwave}), which amounts to solving a Fresnel zone phase retrieval problem. Since the recorded data $I_{ROI}(x,y)$ in a cryoEM system has high level of noise, we prefer to address this problem via a sparse optimization problem. Such an optimization method is preferable over the traditional phase retrieval based on the Gerchberg-Saxton type algorithms \cite{GS1972,Fienup82} which will require us to match the noise in $I_{ROI}(x,y)$ in every iteration. We formulate the retrieval of complex-valued exit wave $g(x,y)$ as a problem of minimizing a cost function of the form:
\begin{align}\label{eq:cost}
    C(g,g^{\ast}) &= C_1 + \alpha \, C_2 \nonumber \\&= || I_{ROI} - |\hat{A} g|^2 ||^2 + \alpha \; L(\nabla g, \nabla g^{\ast}).
\end{align}
Here the operator $\hat{A}$ refers to forward modeling of the exit wave up to the detector and is defined as:
\begin{equation}
    \hat{A} g = \mathcal{F}^{-1} \{H(f_x,f_y) \mathcal{F}[g(x,y)] \},
\end{equation}
with system model $H(f_x,f_y)$ as defined in Eq. (\ref{eq:system_model}). Our aim is therefore to obtain a complex-valued solution $g(x,y)$ whose detector plane irradiance is close to $I_{ROI} (x,y)$. The nature of the solution is additionally controlled by a penalty function $L(\nabla g, \nabla g^{\ast})$ and a positive valued regularization parameter $\alpha$. For penalty term, we use the modified Huber function \cite{khare2015}:
\begin{equation}
  L(\nabla g, \nabla g^{\ast}) = \sum_{k = all \; pixels} \left[\sqrt{1 + \frac{|\nabla g_k|^2}{\delta^2}} - 1\right].   
\end{equation}
In the iterative solution, parameter $\delta$ is made equal to the median of the gradient magnitudes $|\nabla g|$ of the current guess solution. This ensures that at pixels where $|\nabla g_k| >> \delta^2$, the penalty function is similar to the edge preserving total variation (TV) penalty. On the other hand at pixels where $|\nabla g_k| << \delta^2$, $L(\nabla g, \nabla g^{\ast})$ is similar to a quadratic penalty in gradient magnitude which has image smoothing property. The Huber penalty function therefore preserves edges as well as slowly varying grey-scale features in the image solution. It is important to note that the cost function $C(g,g^{\ast})$ is real valued (and positive) while the solution $g(x,y)$ that we are seeking is complex-valued. The descent directions required for the minimization thus need to be defined in terms of the complex or Wirtinger derivatives \cite{brandwood1983complex}:
\begin{equation}
    \nabla_{g^{\ast}} C_1 = - 2 \hat{A}^{\dagger} [(I_{ROI} - |\hat{A}g|^2) (\hat{A}g)],  
\end{equation}
and
\begin{equation}
    \nabla_{g^{\ast}} C_2 = -\nabla \cdot \left[ \frac{\nabla g}{\sqrt{1 + \frac{|\nabla g|^2}{\delta^2}}}\right].
\end{equation}
It is easy to show that the operator $\hat{A}^{\dagger}$ corresponds to application of the complex conjugate transfer function:
\begin{equation}
    H^{\ast}(f_x,f_y) = H_0^{\ast}(f_x,f_y) \exp[i\pi \lambda z \rho^2 - i (\pi/2) C_s \lambda^3 \rho^4],
\end{equation}
to the complex valued field in the detector plane. The expression for $\nabla_{g^{\ast}} C_1$ is essentially what has been used in the Wirtinger flow algorithm \cite{candes2015phase} for the phase retrieval problem. Instead of selecting the parameter $\alpha$ empirically, we use a novel optimization approach - mean gradient descent (MGD) - that allows the solution to move to an optimal solution point from where both $C_1$ and $C_2$ cannot decrease simultaneously \cite{Rajora19}. MGD is inspired by a successful algorithm called as ASD-POCS (adaptive Steepest Descent - Projection onto convex sets) in X-ray computed tomography \cite{Sidky2008}. While the first use of this optimization approach was reported for single-shot interferogram analysis, MGD is a generic methodology that can apply to optimization problems such as the present problem  stated in Eq. (\ref{eq:cost}). The MGD methodology updates the guess solution in the direction that bisects the descent directions corresponding to the two terms $C_1$ and $C_2$ of the cost function. We define two unit vectors:
\begin{equation}
   \mathbf{\hat{u}}_1 = \frac{\nabla_{g^{\ast}} C_1}{|| \nabla_{g^{\ast}} C_1 ||_2}, 
\end{equation}
and
\begin{equation}
   \mathbf{\hat{u}}_2 = \frac{\nabla_{g^{\ast}} C_2}{|| \nabla_{g^{\ast}} C_2 ||_2}, 
\end{equation}
A vector along the direction that bisects the the two directions $\mathbf{\hat{u}}_1$ and $\mathbf{\hat{u}}_2$ may be defined as:
\begin{equation}
    \mathbf{u} = \frac{\mathbf{\hat{u}}_1 + \mathbf{\hat{u}}_2}{2}.
\end{equation}
The $(n+1)$-th MGD iteration then progresses as follows:
\begin{equation}
    g^{(n+1)} = g^{(n)} - \tau \, ||g^{(n)}||_2 \, [\mathbf{u}]_{g = g^{(n)}}.
\end{equation}
The main idea behind MGD is that the change in the solution $g$ due to the individual terms of the cost function during each iteration is made equal in magnitude. The step size $\tau$ was held constant and equal to $10^{-4}$ for simplicity in the illustrations shown in this paper, however, strategies using variable $\tau$ are possible. Since the vector ${\bf u}$ can have norm in the range $[0,1]$, the small step size as used here ensures that the solution changes in norm by small amount in one iteration. For starting guess of this procedure we use a real image with uniformly random pixel values with mean equal to the mean of $\sqrt{I_{ROI}(x,y)}$, where $I_{ROI}(x,y)$ is the cropped part of the micrograph centered on a single particle. This guess has been used since the micrograph is an in-line hologram corresponding to a weak phase object. We observe that starting this random initial guess, the solution progresses in such a manner that the angle between the two vectors $\mathbf{\hat{u}}_1$ and $\mathbf{\hat{u}}_2$ becomes a large obtuse angle. Since the vectors $\mathbf{\hat{u}}_1$ and $\mathbf{\hat{u}}_2$ are complex-valued, in order to define the angle $\theta_{12}$ between $\mathbf{\hat{u}}_1$ and $\mathbf{\hat{u}}_2$, we form two vectors by concatenating the real and imaginary parts of the two unit vectors and define the angle between them via the usual dot product based definition. Thus,
\begin{equation}
    \theta_{12} = \arccos\left(\frac{\mathbf{\hat{v}}_1 \cdot \mathbf{\hat{v}}_2}{||\mathbf{\hat{v}}_1||_2 \, ||\mathbf{\hat{v}}_2||_2}\right),
\end{equation}
where
\begin{align}\label{v1v2}
{\bf v}_1 &= [\textrm{real}(\hat{\bf u}_{1j}) , \textrm{imag}(\hat{\bf u}_{1j})] \nonumber \\
{\bf v}_2 &= [\textrm{real}(\hat{\bf u}_{2j}) , \textrm{imag}(\hat{\bf u}_{2j})].
\end{align}
Here the index $j \,=\, 1, 2, ..., (256)^2$ runs over all the pixels of ROI image matrix. The MGD iteration may be stopped when the angle $\theta_{12}$ exceeds a large obtuse value (e.g. 160 degrees) which indicates that the error and Huber terms of the cost function have essentially achieved a balance. The solution does not appear to change significantly beyond this point for more MGD iterations.
\begin{figure}[tbp]
    \centering
    \includegraphics[width = 0.5\textwidth]{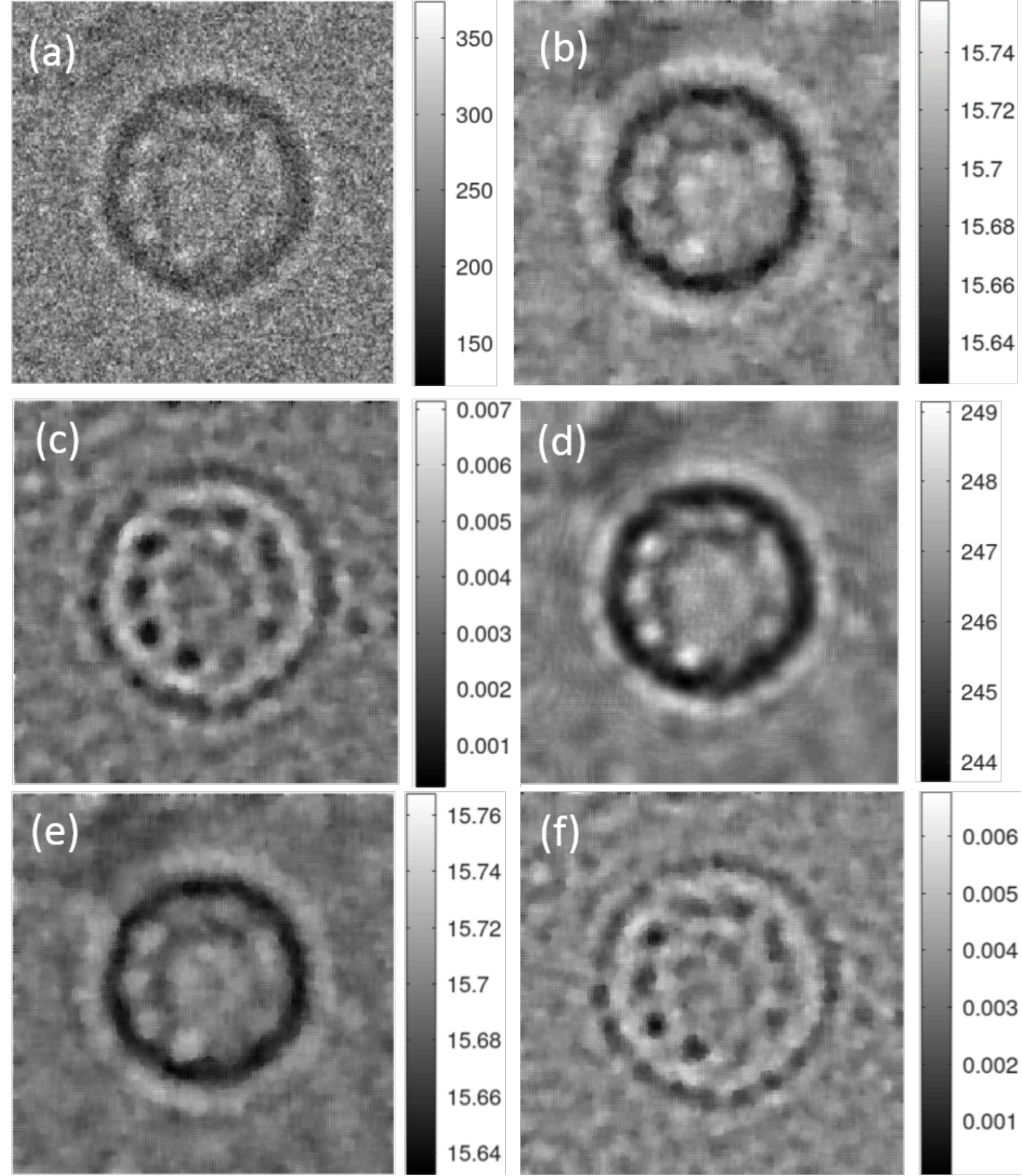}
    \caption{Result of MGD iteration applied to the cropped ROI image. (a) shows the ROI of micrograph used (marked as particle 1 in Fig. \ref{fig1}), (b) and (c) show amplitude and phase associated with the solution $g(x,y)$ after $55$ MGD iterations. This phase retrieval uses the de-focus value of $\Delta z = 3.67 \mu$m as obtained using the peak location of the merit function $M(z)$ as shown in Fig. \ref{fig1}(b). (d) Squared magnitude of the re-projected solution ($|\hat{A}g|^2$) has similar features as in the ROI image but now does not contain noise. (e), (f) Amplitude and phase of the iterative phase retrieval when de-focus value of $\Delta z = 2.7 \mu$m is used.}
    \label{fig3}
\end{figure}
Figure \ref{fig3} shows typical recovered amplitude and phase map for a single particle using the MGD iterative approach. The raw ROI intensity image cropped from the micrograph is shown in Fig. \ref{fig3}(a) which corresponds to the yellow box marked as particle 1 in Fig. \ref{fig1}. Figure \ref{fig3}(b), (c) show the amplitude and phase of the complex valued solution $g(x,y)$ which represents the exit field of the single particle. The range of numerical values of the amplitude image in Fig. \ref{fig3}(b) is small indicating minimal amplitude contrast. The range of phase values in Fig. \ref{fig3}(c) also justifies the small phase object approximation that is usually assumed for the cryoEM samples. The quantitative phase map of the exit field shows stunning structural information regarding the particle, which has been obtained using only the cropped ROI of the micrograph. In order to cross-verify this solution, it is forward projected to obtain the detector domain field $\hat{A}g$. The squared magnitude of this forward projected field $|\hat{A}g|^2$ is shown in Fig. \ref{fig3}(d) which shows features similar to that in the cropped micrograph $I_{ROI}(x,y)$ as shown in Fig. \ref{fig3}(a). The forward projected image $|\hat{A}g|^2$ is observed to have significantly reduced noise compared to the raw data.  The the relative error between the ROI image $I_{ROI}(x,y)$ and intensity of the forward projected solution $g(x,y)$ defined as:
\begin{equation}\label{eq:error}
    E = \frac{|| I_{ROI} - |\hat{A}g|^2||_2}{|| I_{ROI}||_2}
\end{equation}
is approximately equal to $12\%$ for the result in Fig. \ref{fig3}(b), (c). We believe that prior literature on cryo-EM has not explored the possibility of recovering the single-particle quantitative phase information as shown in the illustration above. We carried out a similar iterative recovery of the exit wave assuming the de-focus value of $\Delta z = 2.7 \mu$m which is the estimated de-focus for the whole micrograph. The corresponding amplitude and phase of the recovered complex-valued field are showed in Fig. \ref{fig3}(e), (f) respectively. Once again we note that the quantitative phase of the exit wave in Fig. \ref{fig3}(f) has features similar to that in Fig. \ref{fig3}(c), although a somewhat lower contrast and numerical phase range. The relative error $E$ estimated using the amplitude and phase in Fig. \ref{fig3}(e),(f) is still close to 12 \%. 
\begin{figure}[tbp]
    \centering
    \includegraphics[width = 0.75\textwidth]{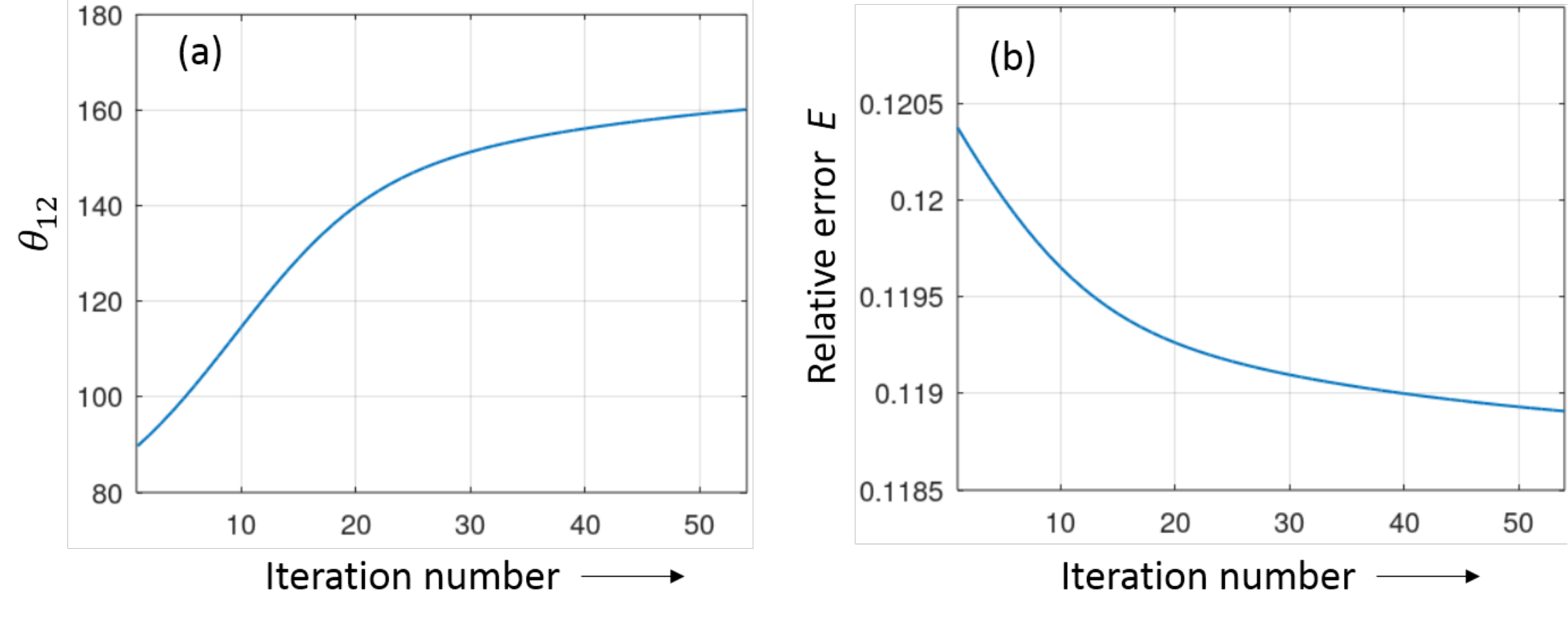}
    \caption{(a)The behavior of angle $\theta_{12}$ between the descent directions $\mathbf{\hat{u}}_1$ and $\mathbf{\hat{u}}_2$ as a function of MGD iterations for the illustration in Fig. \ref{fig3}. The angle is seen to rise steadily to a large obtuse value (in about 90 iterations), indicating that the two terms in the cost function in Eq. (\ref{eq:cost}) have achieved a balance. (b) Behaviour of relative error $E$ defined in Eq. (\ref{eq:error}) with iterations for the illustration shown in Fig. \ref{fig3}.}  
    \label{fig4}
\end{figure}
The behaviour of angle $\theta_{12}$ between the descent directions corresponding to the two terms of the cost function in Eq. (\ref{eq:cost}) as a function of iteration number for recoveries in Fig. \ref{fig3}(b), (c) is shown in Fig. \ref{fig4}. We observe that the angle $\theta_{12}$ rises to above $160$-degrees in $54$ iterations. The relative error $E$ as a function of iterations is also plotted in Fig. \ref{fig4}(b).
\begin{figure}[tbp]
    \centering
    \includegraphics[width = 0.75\textwidth]{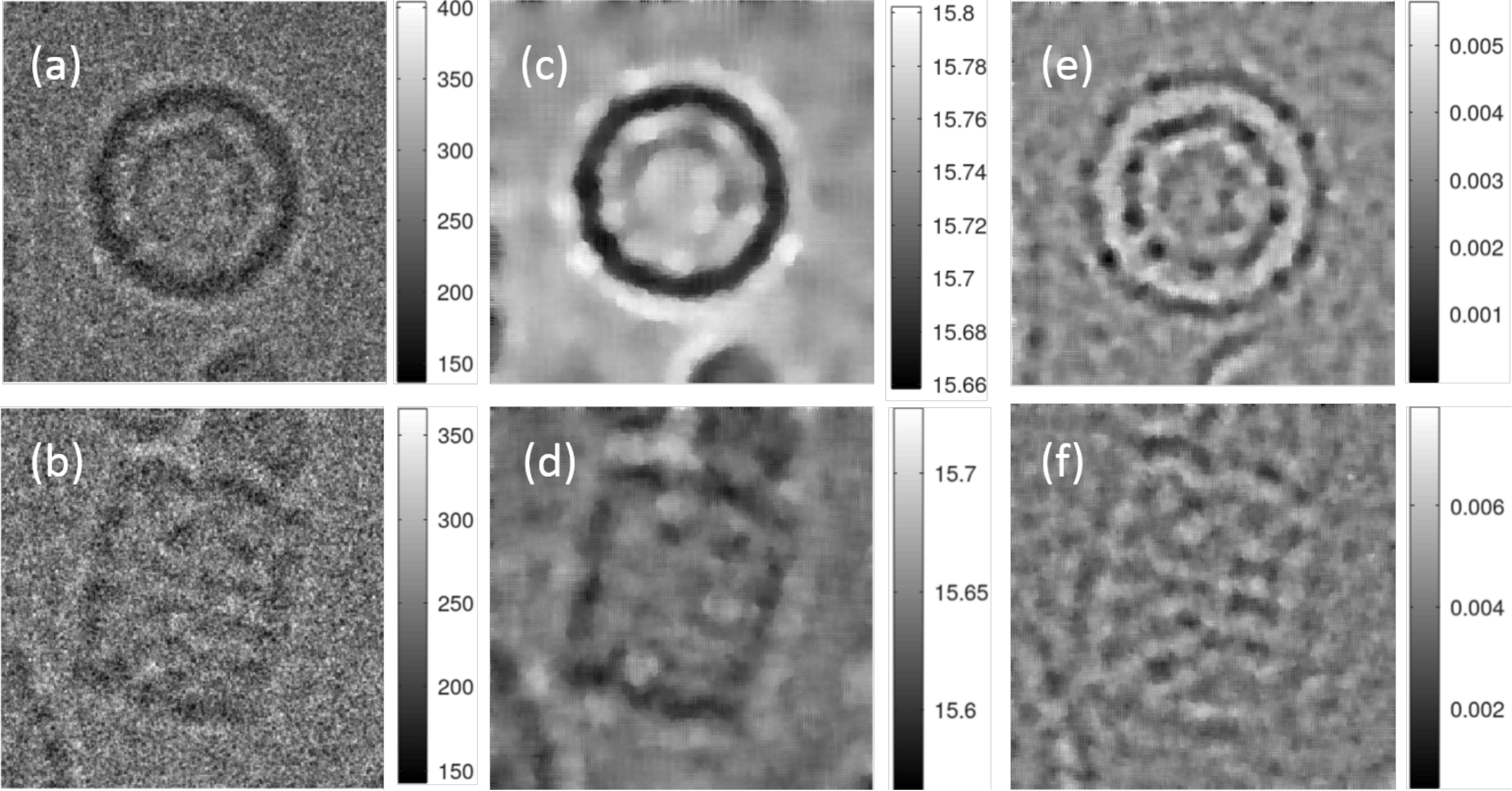}
    \caption{Exit wave recovery for two other particles (marked as 2, 3 in the micrograph shown in Fig. \ref{fig1}). (a), (b) raw cropped ROIs from the micrographs, (c), (d) reconstructed amplitudes, (e),(f) reconstructed phase images.}
    \label{fig5}
\end{figure}
Figure \ref{fig5} shows exit wave recoveries for two other particles (marked as 2, 3 in the micrograph shown in Fig. \ref{fig1}). The cropped ROI raw data (Fig. \ref{fig5}(a),(b)), the recovered amplitude (Fig. \ref{fig5}(c),(d)) and phase (Fig. \ref{fig5}(e),(f)) are displayed in this illustration. The de-focus values estimated for these two particles using the merit function peak are $2.47 \mu$m and $3.395 \mu$m respectively. Once again the amplitudes $|g(x,y)|$ are nearly flat and the quantitative phase maps show interesting structural information relevant to the single particles in different orientations. The relative error $E$ for these two cases defined as per Eq. (\ref{eq:error}) is once again approximately 12\%.

\section{Quantitative phase retrieval for single particles in the Apoferritin data}
\begin{figure}[tbp]
    \centering
    \includegraphics[width = 0.5\textwidth]{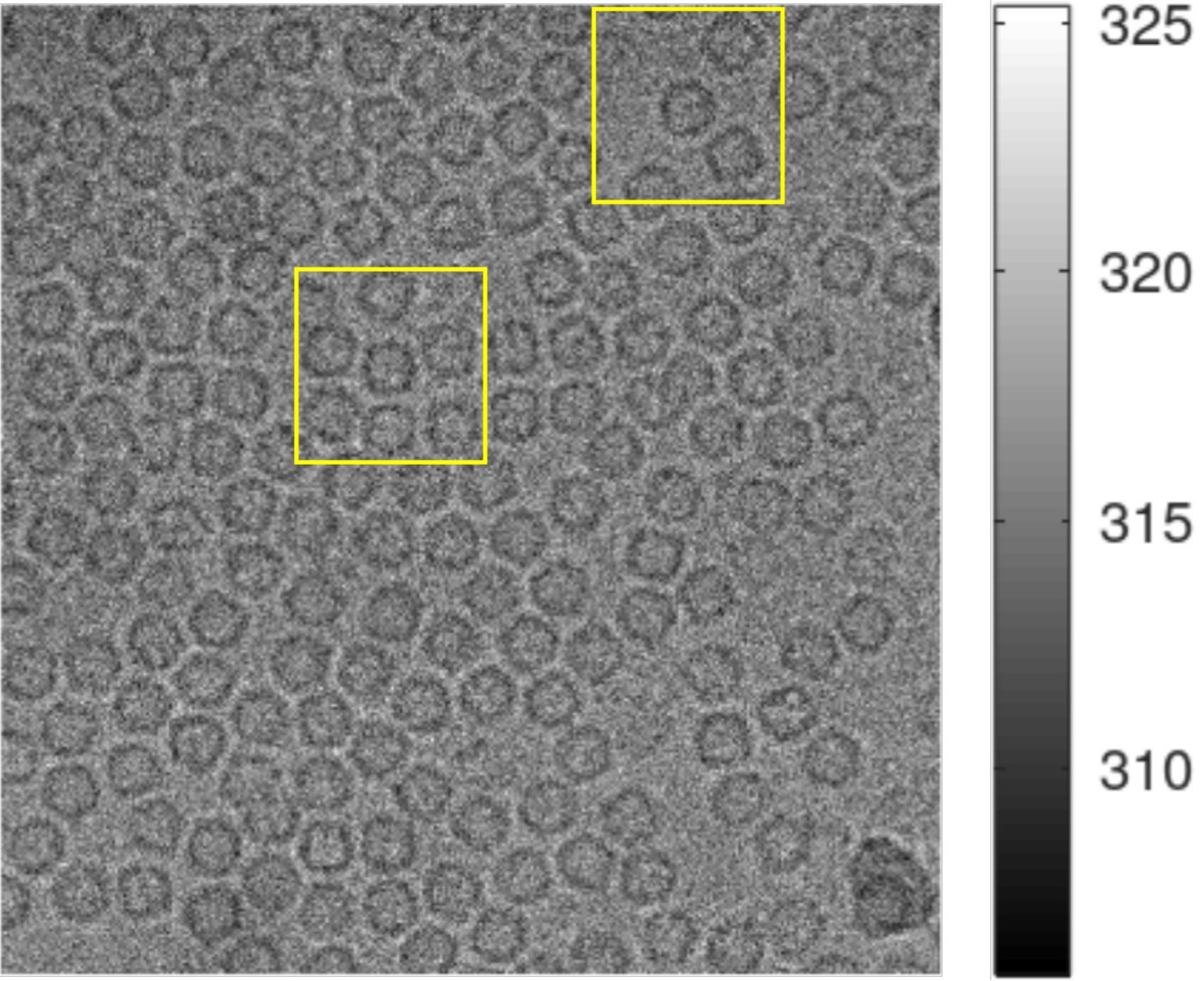}
    \caption{A motion-corrected micrograph from Apoferritin data EMPIAR-10146. The micrograph has a de-focus value of 2.5 $\mu$m estimated using CTFFIND4. The pixel size is $1.5$ Angstrom and the image has $1240 \times 1200$ pixels.}
    \label{fig6}
\end{figure}

\begin{figure}[htbp]
    \centering
    \includegraphics[width = 0.75\textwidth]{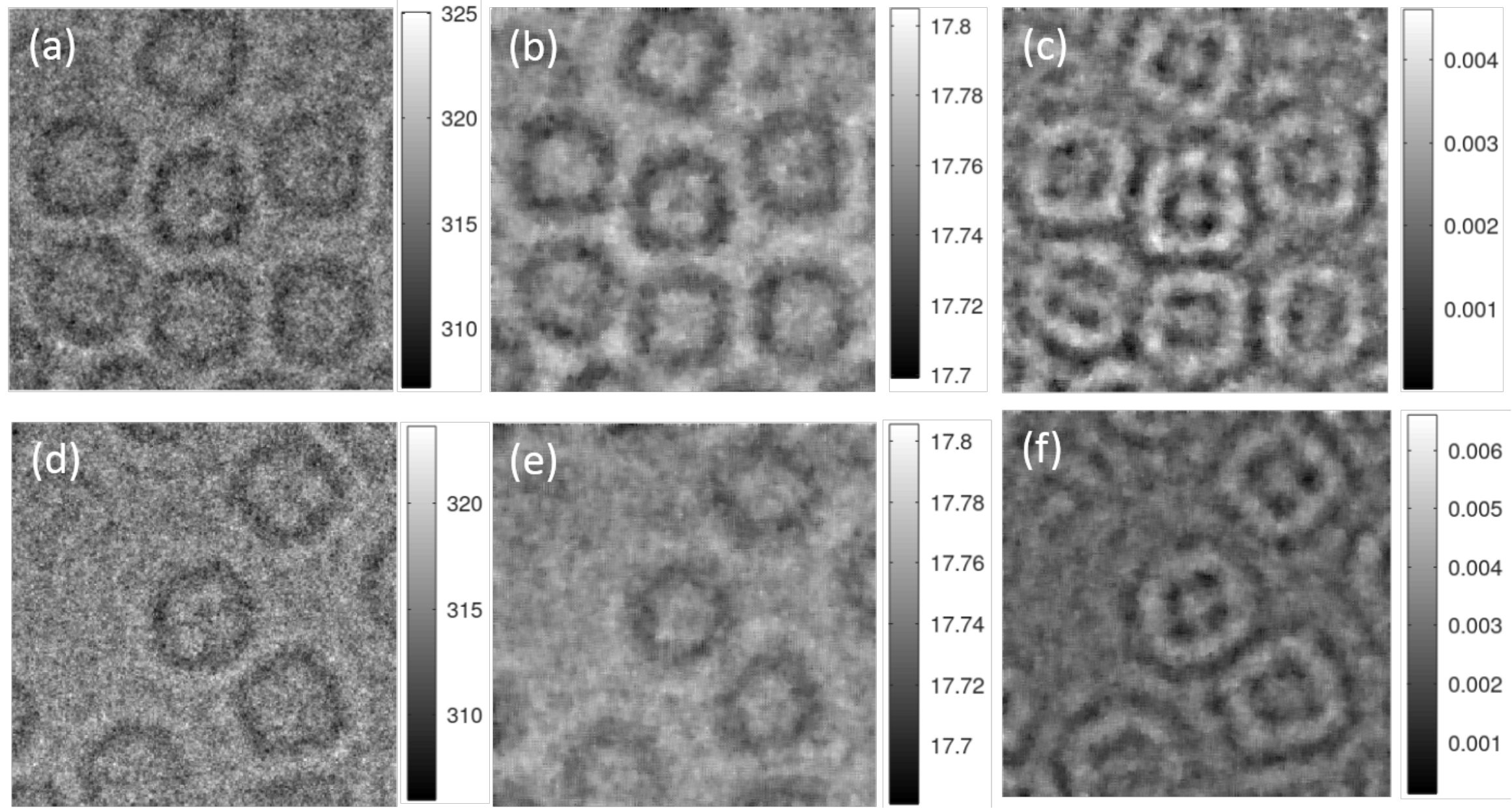}
    \caption{Iterative recoveries for two regions in the Apoferritin micrograph. (a),(d): Raw ROI regions cropped from micrograph, (b),(e): amplitude and (c),(f): phase of recovered exit wave.}
    \label{fig7}
\end{figure}

In order to emphasize the generality of the Fresnel-zone phase retrieval approach discussed above, we apply identical focus plane detection and the MGD iterative procedure for two ROI regions from a micrograph in the Apoferritin data (EMPIAR-10146). The optimization problem described by the cost function in Eq. (\ref{eq:cost}) may also be solved using more traditional approaches, but this will involve empirical tuning of the parameter $\alpha$ depending on the detector noise. The MGD approach however does not need any empirical tuning of free parameters. The Apoferritin data has been recorded on a 300 kV system with a direct electron detector and spherical aberration corrector. The effective pixel size in the micrographs for this data is $1.5$ Angstrom. The spherical aberration coefficient $C_s$ for this system is negligible. A motion corrected micrograph from this data is shown in Fig. \ref{fig6}. The estimated de-focus value for this micrograph determined using CTFFIND4 is approximately 2.5 $\mu$m and therefore we search for the focus plane in the range $z_1 = 1.5 \,\mu$m to $z_2 =3.5 \,\mu$m. Our focus detection procedure based on the merit function peak however led to de-focus values of $\Delta z = 3.4 \mu$m and $3.2 \mu$m that we have used for the iterative procedure. We have used two ROI regions of the size 256 $\times$ 256 pixels (marked in Fig. \ref{fig6}) for phase recovery. 
The complex exit wave recovery for the two ROIs shown in Fig. \ref{fig6} are illustrated in Fig. \ref{fig7}. Once again the amplitude of the exit is nearly constant over the image area and the quantitative phase maps show distinct structural features corresponding to the single particles. Interestingly, the relative error performance of these recoveries as measured by the error function $E$ defined in Eq.(\ref{eq:error}) is $0.5 \%$. Since the micrograph has been recorded using a direct electron detector with lower detector noise compared to the older KLH data, a better error performance for the MGD based recoveries is as expected.  
\begin{figure}[htbp]
    \centering
    \includegraphics[width = 0.5\textwidth]{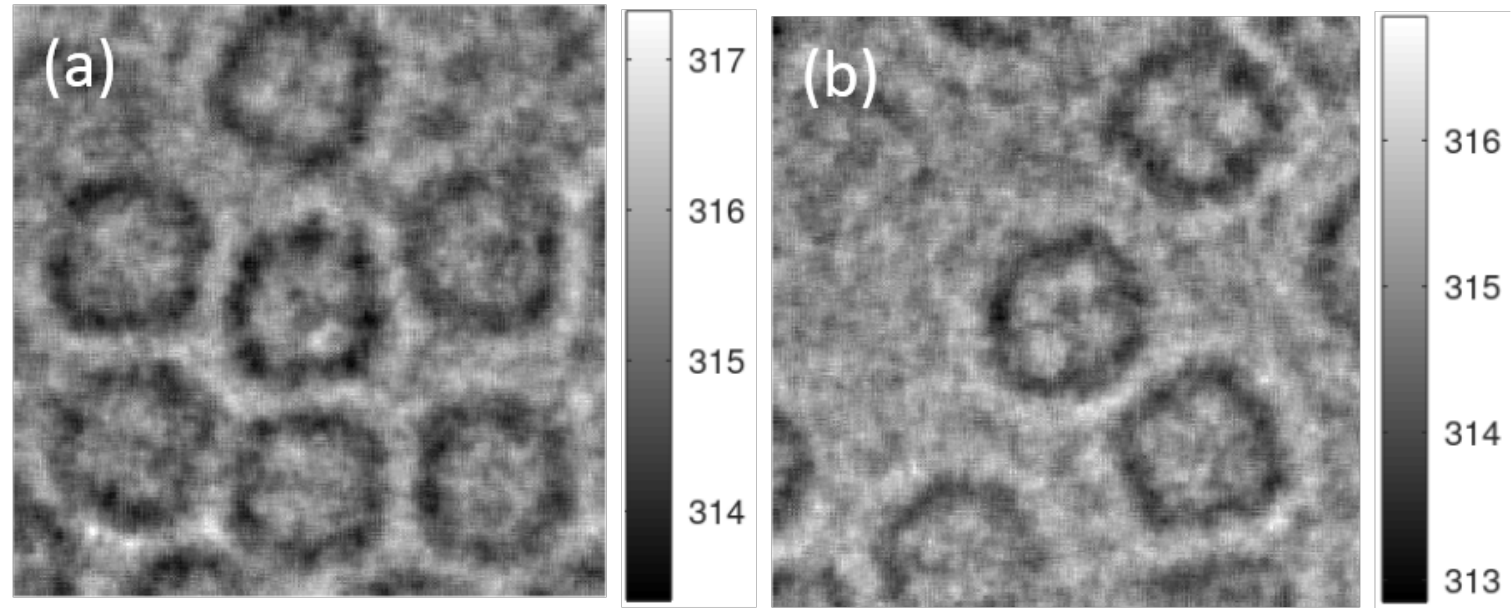}
    \caption{ Squared magnitude of the re-projected solution ($|\hat{A}g|^2$) for the two phase recoveries shown in Fig. \ref{fig7}. The relative error $E$ for this dataset is $0.5\%$.}
    \label{fig8}
\end{figure}
The intensity maps corresponding to the forward projected solutions given by $|\hat{A}g|^2$ for the two cases in Fig. \ref{fig7} are shown in Fig. \ref{fig8}(a), (b) respectively and show features similar to those in Fig. \ref{fig7}(a), (d) with reduced noise. The results suggest that the methodology for focus plane detection using the sparsity of gradient merit function and the iterative procedure for complex-valued exit wave recovery applies uniformly to older CCD detector based systems as well as the newer systems using DED detector.  

\section{Discussion and future work}
We have described a methodology for recovering physically meaningful quantitative phase maps associated with single particles information from cryoEM micrographs. In particular by treating the cryoEM micrograph as an in-line hologram, we use a sparsity of gradient merit function for estimating defocus associated with each particle using only the local ROI information surrounding a single particle. While this methodology for defocus estimation needs further study, the initial results indicate that it may be valuable for the cryoEM community. Further, using this defocus value, an iterative approach for sparsity assisted Fresnel-zone phase retrieval (including the effect of spherical aberration) is presented. The methodology provides quantitative phase information for individual single-particles in their specific orientations. It is observed that the recovered complex-valued exit wave is consistent with the raw detector data as permitted by the noise in the detected micrograph. As shown in our illustrative results, the overall procedure is uniformly applicable to micrograph data recorded using older CCD detector based systems as well as newer systems employing the direct electron detectors. The present processing steps in the 3D reconstruction chain in cryoEM employs 
steps such as contrast inversion and normalization of particle images, as a result the 3D reconstruction map is essentially qualitative in nature and each voxel of the map cannot be claimed to represent any physical quantity in a strict sense. The possibility of recovering meaningful quantitative phase information for individual particles is valuable in our opinion, as the phase map of the exit wave is related to the tomographic projection of the individual particle in given orientation. The implications of the quantitative phase information in 3D reconstruction remains to be explored in future.

\section*{Funding}
AP acknowledges support from Prime Minister's Research Fellowship. MB acknowledges partial support from Department of Biotechnology (Award number: BT/ PR29264/BRB/10/1710/2018). KK acknowledges partial support from Department of Science and Technology India (Award number: ID/MED/34/2016). 

\bibliography{references}

\end{document}